\documentclass[floatfix,twocolumn,rmp,superscriptaddress]{revtex4}

\usepackage{graphicx,amsmath,amssymb}

\begin{document}

\title{Evolution of Rogue Waves in Interacting Wave Systems}

\author{Andreas Gr\"onlund}
\affiliation{Department of Mathematics, Uppsala University, SE--751 06 Uppsala, Sweden}
\affiliation{Ume{\aa} Plant Science Center, Department of Forest Genetics and Plant Physiology, Swedish University of Agricultural Sciences, SE--901 83 Ume{\aa}, Sweden}

\author{Bengt Eliasson}
 \affiliation{Department of Physics, Ume\aa\ University, SE--901 87 Ume\aa, Sweden}

\author{Mattias Marklund}
\affiliation{Department of Physics, Ume\aa\ University, SE--901 87 Ume\aa, Sweden}

\begin{abstract}
Large amplitude water waves on deep water has long been known in the sea faring community, and the cause of great concern for, e.g., oil platform constructions. The concept of such freak waves is nowadays, thanks to satellite and radar measurements, well established within the scientific community. There are a number of important models and approaches for the theoretical description of such waves. By analyzing the scaling behavior of freak wave formation in a model of two interacting waves, described by two coupled nonlinear Schr\"{o}dinger equations, we show that there are two different dynamical scaling behaviors above and below a critical angle $\theta_c$ of the direction of the interacting waves  below $\theta_c$ all wave systems evolve and display statistics similar to a wave system of non-interacting waves. The results equally apply to other systems described by the nonlinear Schr\"{o}dinger equations, and should be of interest when designing optical wave guides.
\end{abstract}

\maketitle

\section{Introduction}
The occurrence of rogue waves, i.e  waves that are at least twice the size of the significant wave height of the surrounding waves, has long been a well known and much feared subject among professionals in the off-shore and sea faring business~\cite{smith,dean,kharif-pelinovsky}.  The first truly scientific measurement of such a rogue wave was done at the Draupner oil platform in the North Sea off the coast of Norway on January 1, 1995
\cite{Trulsen97}, where the so-called Draupner wave or New Years wave was observed. Since then, a large number of measurements, using different techniques, have been made in areas prone to such rogue waves, and the concept of rogue or freak waves is now scientifically well established~\cite{lavrenov,mori-etal,liu-pinho,kharif-pelinovsky,divinsky-etal,stansell}. The findings show clearly that the rogue wave phenomenon is far more common than predicted by linear wave models. Both shallow and deep water wave behavior can to some degree be captured by simple nonlinear models, e.g.\ the Korteweg-de Vries or nonlinear Schr\"odinger equations (NLSE) \cite{Benney69,Toffoli}. For deep water waves, modified NLSE models, such as the Dysthe equation~\cite{dysthe}, have been proposed to more accurately characterize such waves. Simulations of the exact dynamical equations have shown the formation of freak waves in the nonlinear stage of the Stokes wave modulational instability \cite{Zakharov}. Some of the main features of rogue waves are nevertheless still elusive in the simplified model descriptions. Hammack et al. \cite{Hammack} studied the nonlinear interaction between large amplitude water waves in the laboratory and observed a dependence of the modulational instability on the angle between two large amplitude water waves. Recently, Onorato \textit{et al}~\cite{onorato-etal2} showed that interesting behavior arose if freak waves were modeled using nonlinearly interacting water waves. This model was later shown to give rise to wave pattern close to present rogue wave observations~\cite{shukla-etal}.

\section{Model}

In this paper we present properties of a model of nonlinearly interacting water waves and investigate to which degree the interaction promotes extreme waves, both in terms of the frequency at which waves of different height are occurring, and of the wave system's short and long term behavior. We will focus on a model for rogue waves based on two coupled NLSE~\cite{Hammack,onorato-etal2,shukla-etal}, which permits large-scale simulations needed to bridge the gap to real-world wave statistics. The model system consists of two interacting waves $A$ and $B$, both traveling into a region of interaction. The wave system and its time evolution is illustrated in Fig.~1a.

\begin{figure*}[t]
\center
\resizebox*{\linewidth}{!}{\includegraphics{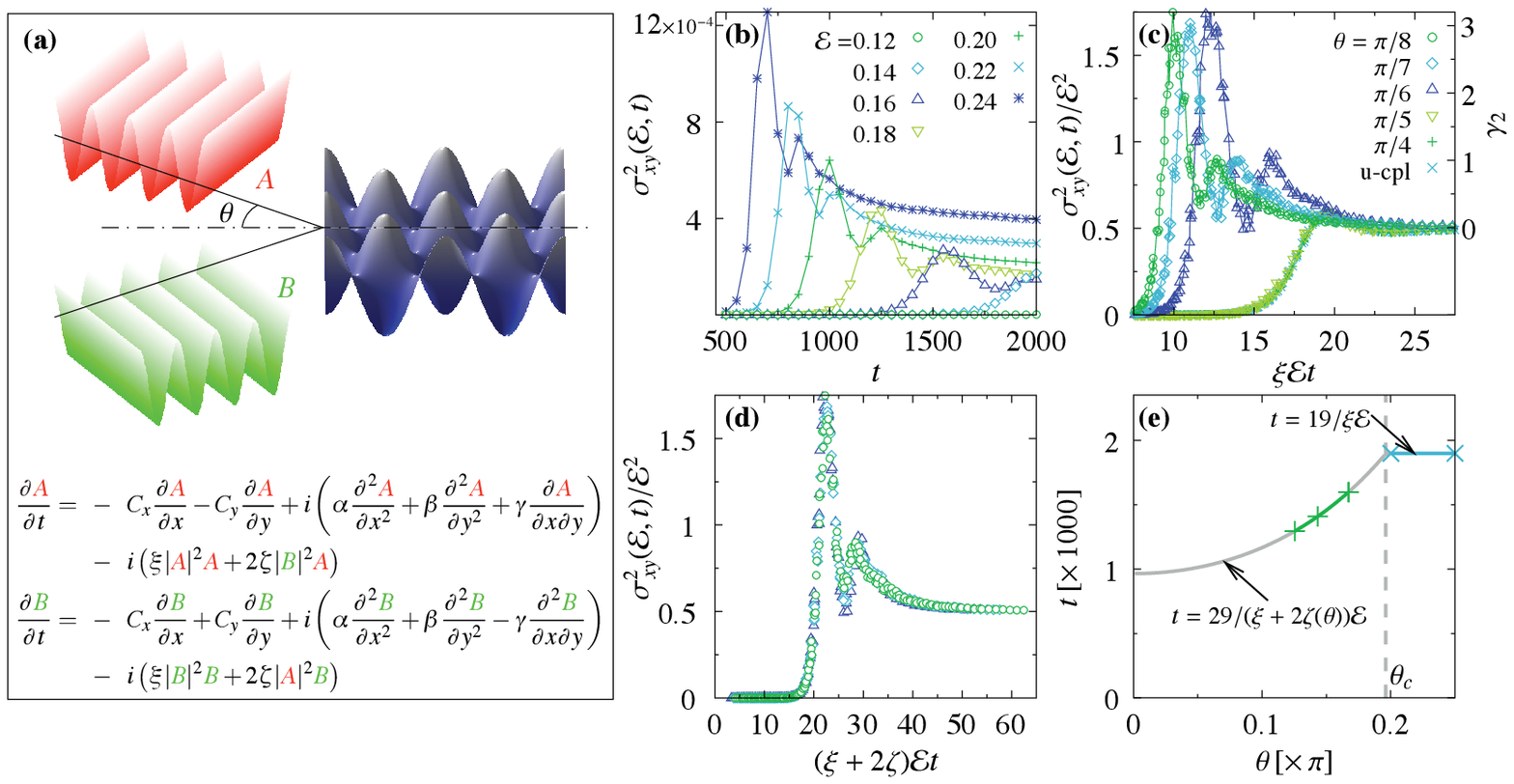}}
\caption{ (a) An illustration of the wave system and the NLSE. Two waves $A$ and $B$ with the angle $\theta$ relative the dichotome are interacting and give collective wave dynamics in the blue region. The coupling constant $\zeta=\zeta(\theta)$ dictates the level of interaction between the two waves $A$ and $B$.  (b) Time evolution of the variance for $\theta=\pi/8$ and different values on ${\cal E}$. (c) a collapse of $\sigma_{xy}$, the energy ${\cal E}$, the nonlinear coefficient $\xi$ and time $t$ for different values on $\theta$ and for the uncoupled (u-cpl) system. The right-hand vertical axis shows the kurtosis $\gamma_2$ of wave surface $\eta(x,y)$.(d) A collapse for $\theta=\pi/8,\pi/7$ and $\pi/6$ as in (c) but including the coupling coefficient as $(\xi+2\zeta){\cal E}$. (e) The different scaling behavior above and below $\theta_{\rm c}$. $t$ is the time to reach the second maxima of $\sigma_{xy}^2$ for systems below $\theta_{\rm c}$ and the time to reach the only maximum of $\sigma_{xy}^2$ for systems above $\theta_{\rm c}$. The green plus signs (+) are, from left to right, for $\theta=\pi/8,\pi/7$ and $\pi/6$. The blue, crosses ($\times$), are for $\theta=\pi/5$ and $\pi/4$. All with ${\cal E}=0.2$. The gray line shows calculated times for different $\theta$s using the scaling behavior in Fig.~1c and d. In all cases, the errors are smaller than symbols and the number of grid points $N_x\times N_y=256\times 256$ is used in the $x$ and $y$ direction.}
\label{Fig1}
\end{figure*}

We analyze the statistical and spectral properties of a system of two nonlinearly interacting water waves by performing extensive numerical simulations.   All measured quantities are obtained from multiple realizations, each starting with different initial wave envelopes $A$ and $B$ by giving them a small amplitude, noise term. The evolution of the wave system is calculated from the slowly varying wave envelopes  $A$ and $B$, which are related to the actual wave surface $\eta$ as $\eta=\eta_A+\eta_B$ with $\eta_{A}=(1/2)A(\mathbf{r},t)\exp{\left(ik_xx + ik_yy-i\omega t \right)} +$ c.c, and $\eta_{B}=(1/2)B(\mathbf{r},t)\exp{\left(ik_xx - ik_yy - i\omega t \right)} +$ c.c., where c.c. denotes the complex conjugate, $k_x$ and $k_y$ are the $x$ and $y$ components of the wave vector, and $\omega$ is the wave frequency. The latter is related to the wave vector via the water wave dispersion relation
$\omega=\sqrt{g\kappa}$, where $g$ is the gravitational acceleration and $\kappa=\sqrt{k_x^2+k_y^2}$ is the magnitude of the wave vector. The time evolution of the wave envelopes are given by the coupled NLSEs, displayed in Fig.~1a. As we can see the last term dictates the coupling between the two waves with the coupling constant $\zeta=\zeta(\theta)$ which ultimately is given from the angle $\theta$.  By setting $\zeta=0$ we obtain an uncoupled wave system.

In short we have \cite{Hammack,onorato-etal2,shukla-etal}: the group velocity components $C_{x,y} = \omega k_{x,y}/2\kappa^2$, the group dispersion coefficients $\alpha = \omega\left(2k_y^2-k_x^2\right)/8\kappa^4$, $\beta=\omega\left(2k_x^2-k_y^2\right)/8\kappa^4$ and $\gamma = -3\omega k_xk_y/4\kappa^4$, the nonlinear coefficient $\xi = \omega\kappa^2/2$, and finally, the coupling coefficient
$\zeta =  \omega\left(k_x^5 - k_x^3k_y^2 - 3k_xk_y^4 - 2k_x^4\kappa + 2k_x^2k_y^2\kappa + 2k_y^4\kappa\right)$ $/ 2\kappa^2\left(k_x - 2\kappa\right)$. The vector components $k_x$ and $k_y$ can be expressed from the angle between the wave vectors and the dichotome as $k_x=\kappa\cos{\theta}$ and $k_y=\kappa\sin{\theta}$. We have normalized the spatial coordinates and wave envelopes as ${\bf r}\mapsto \kappa^{-1} {\bf r}'$, $A\mapsto \kappa^{-1} A'$, and $B\mapsto \kappa^{-1} B'$, and time as $t\mapsto \omega^{-1} t'$ to obtain dimensionless coupled NLSEs in terms of the dimensionless, primed variables, and where $\kappa$ and $\omega$ are eliminated. For convenience, we have omitted the primes below.

We simulate the dynamics of the two interacting waves by using a pseudo-spectral routine---calculating the spatial derivatives in the spectral domain with periodic boundary conditions, and the fourth-order Runge-Kutta algorithm as time integrator. We use the grid sizes $\Delta x=\Delta y=1$ and the time step $\Delta t=0.5$. The number of grid points are $N_x\times N_y=256\times 256$ in Fig. 1 and $N_x\times N_y=128\times128$ in Fig. 2. As initial conditions, we set the $A$ and $B$ to constant amplitudes $A_0$ and $B_0$, with $A_0=B_0$ having values in the range 0.05--0.12. To seed the instability and generate multiple realizations for each parameter setting we add a small amplitude noise to $A$ and $B$ of the order $A_0/100$ and $B_0/100$.

\section{Results}

The total energy $E = \int \left( \vert A(x,y,t)\vert^2 + \vert B(x,y,t)\vert^2 \right) \,dx\,dy$ is a conserved quantity. Locally the energy fluctuates, and the wave energy fluctuations of the system can be quantified by the variance $\sigma^2_{xy}({\cal E},t) = \frac{1}{L^2}\int \left(\vert A(x,y,t)\vert^2 + \vert B(x,y,t)\vert^2 - {\cal E}\right)^2\,dx\,dy$, where ${\cal E}=E/L^2$ is the average energy density.

In Fig.~1b we display the fluctuations $\sigma^2_{xy}({\cal E},t)$ for a system of two interacting waves $A$ and $B$ measured at different times $t$ and with different energies ${\cal E}$ using the angle $\theta=\pi/8$. We see that $\sigma^2_{xy}({\cal E},t)$ initially increases with time for all ${\cal E}$ and reaches a maximal value, then decreases and eventually stabilizes at a value that is larger for larger values of ${\cal E}$. Moreover, Fig.~1b also indicates that $\sigma^2_{xy}({\cal E},t)$ can be described by a function of only one variable. Specifically, we assume that the shape of $\sigma^2_{xy}({\cal E},t)$ can be expressed in terms of a dimensionless scaling function  of only one variable, $F_{\sigma}(X)$. We write this as $\sigma^2_{xy}({\cal E},t) = {\cal E}^a\,F_{\sigma}((\kappa^2 {\cal E})^b (\omega t)^c)$. For dimensional reasons we then expect $a=2$, as $F_{\sigma}$ is dimensionless, but leave this for the collapse plot.

Fig.~1c shows a collapse plot of $\theta=\pi/8\rightarrow\pi/4$ and an uncoupled (u-cpl) system (one collapsed curve for each case), with the values $a=2$, $b=1$ and $c=1$. Hence, we can read out a scaling law of the form $\sigma^2_{xy}({\cal E},t) = {\cal E}^2\,F_{\sigma}(\xi {\cal E} t)$, since $\xi = \omega \kappa^2/2$ --- the nonlinear coefficient. We also see that as $X \rightarrow \infty$, $F_{\sigma}(X)$ approaches the constant value $0.5$, which yields the property  $\sigma^2_{xy}({\cal E},t) \sim 0.5 {\cal E}^2$ as $t\rightarrow \infty$. The scaling function tells us that the time evolution of the wave system display the same behaviour and is only scaled by the wave energy (and the nonlinear coefficient). The right-hand vertical axis in Fig~1c show the positive values of the excess kurtosis of the wave surface $\gamma_2=\mu_4(\eta)/\sigma(\eta)^2-3$, as this can be mapped onto our dimensionless scaling function since the variance of ${\cal E}$ is proportional to the fourth moment of the wave surface and ${\cal E}$ is proportional to the variance of the wave surface. This only holds when the fluctuations of the envelopes are large, and therefore we only display the the mapping for $\gamma_2>0$, which are obtained from processes with a wider-than-normal-distribution. We see that the coupled wave system reaches a maximum kurtosis of 3 and that all wave systems finally approach zero, the kurtosis of a normal distribution. Similar results has been obtained for simulations of shallow water waves in crossing seas \cite{Toffoli}.

Note that for larger angles $\theta=\pi/5,\pi/4$ and uncoupled wave systems the curves fall on top of each other, in contrast to $\theta=\pi/8,\pi/7$ and $\pi/6$ which are shifted. It was noted in Refs.~\cite{onorato-etal2,shukla-etal} that there is a critical angle $\theta_{\rm c}={\rm arctan}(1/\sqrt{2})\approx 0.615\,{\rm rad}\approx 35.3^\circ$ at which the interaction changes character, making systems with $\theta<\theta_{\rm c}$ more unstable due to overlapping instability regions and forming rogue waves at shorter time scales. This may explain the difference between $\theta=\pi/8,\pi/7,\pi/6<\theta_{\rm c}$ where the waves interact strongly, and $\theta=\pi/5,\pi/4>\theta_{\rm c}$ where the waves behave as uncoupled (or single) wave systems. 

The space-independent harmonic solution is given as~\cite{shukla-etal}, $A_{eq}=B_0\exp[-i(\xi B_0^2 + 2\zeta A_0^2)t]$ and $A_{eq}=B_0\exp[-i(\xi B_0^2 + 2\zeta A_0^2)t]$, where $A_0$ and $B_0$ are the initial, average, values of the wave envelopes. Since $A_0=B_0$, the average energy density is ${\cal E}=(A_0^2+B_0^2)=2A_0^2$, and we see that $A_{eq}=A_0\exp[i(\xi+2\zeta)A_0^2t]=A_0\exp[(i/2)(\xi+2\zeta){\cal E}t]$. So in the region of overlapping instabilities of the two waves, the time evolution might be scaled by  $(\xi+2\zeta)$ rather than just $\xi$. We test this as before by performing a collapse for $\theta=\pi/8,\pi/7$ and $\pi/6$. In Fig.~2d we see that all three curves fall onto one and we can read out a new scaling as  $\sigma^2_{xy}({\cal E},t) = {\cal E}^2\,F_{\sigma}[(\xi+2\zeta){\cal E}t]$.

The coupled wave systems with $\theta=\pi/8,\pi/7$ and $\pi/6$ have two local maxima, with the second maxima close to $(\xi+2\zeta){\cal E}t \approx 29$. For systems of waves with no region of overlapping instability, when $35.3^\circ < \theta < 68.02^\circ$, there is only one peak of $F_{\sigma}$, indicating that overlapping instability regions are needed to form the first peak. For $\theta=\pi/5$, $\pi/4$ and uncoupled waves, the only maximum of $\sigma_{xy}$ is found at $\xi {\cal E}t \approx 19$. Our assumption is that the second maxima for $\theta=\pi/8,\pi/7$ and $\pi/6$ and the only maximum for $\theta=\pi/5$ and $\pi/4$ are evolved by additional instabilities seen in both systems and represent the same point in their development, but is reached at different times du to a different scaling.

By plotting the time it takes to reach the second maxima for systems of overlapping instabilities and the only maximum for systems of no overlapping instabilities for different $\theta$s, the two curves should then if our assumption holds intersect at $\theta_c$ and demonstrate that that the different scaling behaviour is separated by $\theta_c$. In Fig.~2e we plot $t$ for the two different scaling behaviors and additionally show the measured values for the different $\theta$s. We indeed see that the two curves intersect at the critical angle $\theta_{\rm c}\approx 0.196\pi \,\,{\rm rad}$ as anticipated. This, together with Fig.~1c and Fig.~1d tells us that
\begin{equation}
  \sigma^2_{xy}(\theta,{\cal E},t) = \left\{\begin{array}{ll} &  {\cal E}^2\,F_{\sigma}\left[\xi {\cal E}t\right] \mbox{if $\theta \geq \theta_c$}\\
          & {\cal E}^2\,F_{\sigma}\left[(\xi+2\zeta(\theta)){\cal E}t\right] \mbox{if $\theta<\theta_c$}\end{array}\right..
\end{equation}
It was also noted in Refs. \cite{onorato-etal2,shukla-etal} that counter-propagating waves and waves propagating at large angles, $68.02^\circ<\theta\leq 90^\circ$ give rise to a new instability perpendicular to their direction of propagation. However, this large angle regime is outside the scope of the present article and is left for future studies.
~
\begin{figure*}[t]
\center
\resizebox*{\linewidth}{!}{\includegraphics{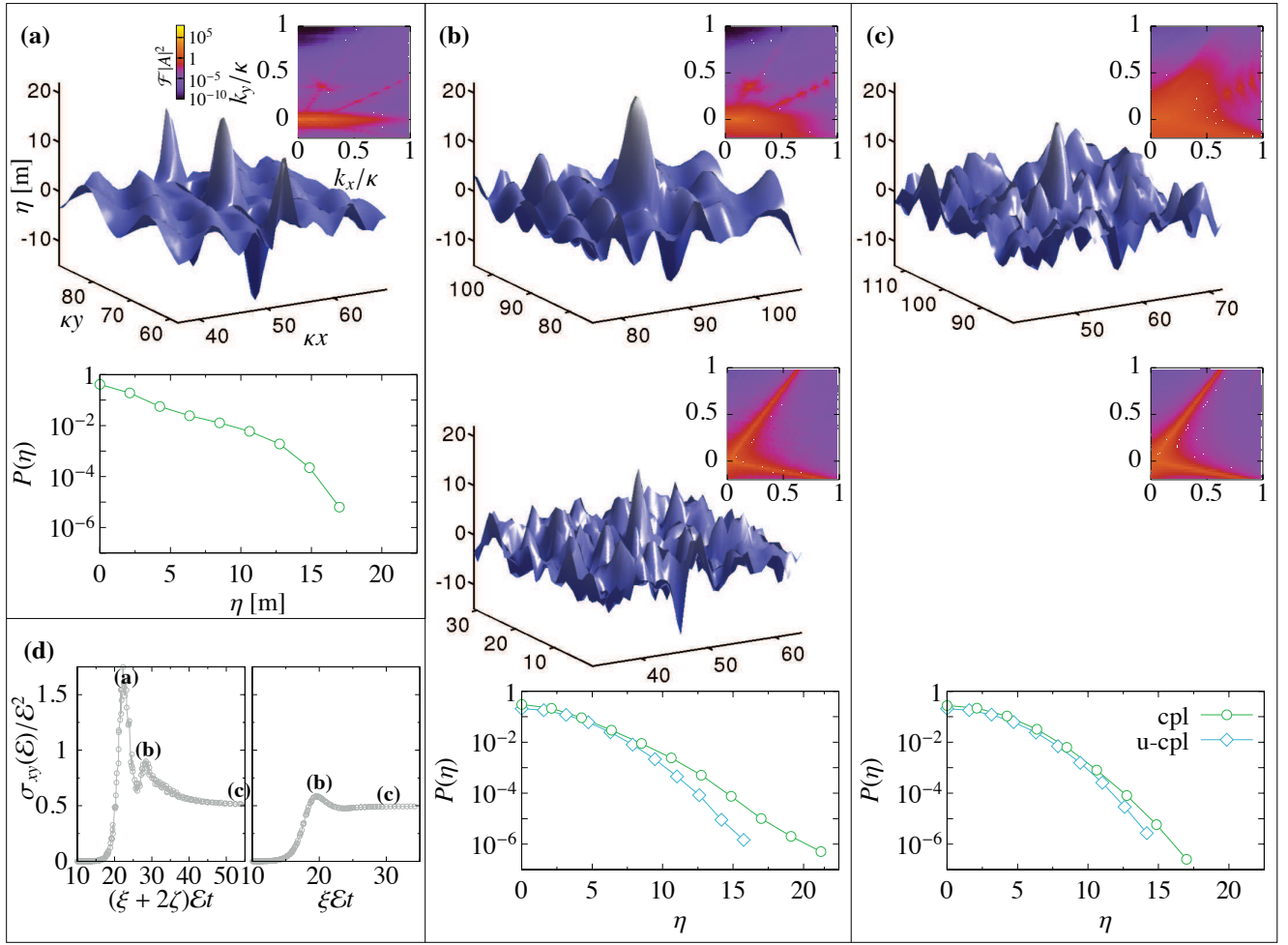}}
\caption{In (a) wave data from snapshots of a coupled system at the time point giving the maximum kurtosis ($\approx 3.2$). Top figure shows the surface elevation $\eta$ (in meters) of the largest occurring wave from all realizations and their averaged power spectrum. Bottom panel shows the surface elevation probability density. In (b) the point in time giving the largest waves in the coupled (cpl) and the uncoupled (u-cpl) system; $\sigma_{xy}$, $(\xi+2\zeta){\cal E} t \approx 30$ and $\xi {\cal E}t\approx 20$, respectively. From top to bottom; the largest wave and a power spectrum of a coupled system, similarly fo the uncoupled system, and finally, their surface elevation probability densities. (c) the same as (b) but at a later time, showing the asymptotic properties of the surface elevations for the two systems. $\sigma_{xy}$, $(\xi+2\zeta){\cal E} t \approx 55$ for the coupled system and $\xi {\cal E}t\approx 30$ for the uncoupled system. In (d) an illustration of where the snapshots are taken. We in all cases generate multiple realizations using ${\cal E}=0.2$, $\theta=\pi/8$ and the number of grid points $N_x\times N_y=128\times 128$.}
\label{Fig2}
\end{figure*}

Typical data from ocean waves give~\cite{hasselmann}, the wave frequency 0.09 Hz, $\omega=0.56 \textnormal{s}^{-1}$, and $\kappa = 0.033\textnormal{m}^{-1}$. We set $A = B = 0.1/\kappa = 3 \textnormal{m}$. In Fig.~2a we display the wave surface $\eta$, the power spectrum ${\cal F}\vert A \vert^2$, and the surface elevation probability density function $P(\eta)$ of an interacting wave system with $\theta=\pi/8$. Note that only the positive part of the surface elevation probability density is displayed, as the distribution is symmetric around zero. The snapshots are taken where the largest kurtosis is found (See Fig.~1c). This is exactly at the end of the exponential growth regime, with the wave envelopes $A$ and $B$ forming waves fronts extending in the $y$ direction, traveling in the $x$ direction. In Fig.~2b we look at the second peak of $\sigma_{xy}$, $(\xi+2\zeta){\cal E} t \approx 30$, of the coupled wave (upper), and compare it with the $\sigma_{xy}$-maxima, $\xi {\cal E}t\approx 20$, of an uncoupled wave system (middle), also with $\theta=\pi/8$. This is where the two systems are showing the largest waves, despite the fact that the coupled system has a larger kurtosis in Fig~2a. The largest wave found in the coupled system is 21.24m, significantly larger than in the uncoupled case, 15.75m. The distribution $P(\eta)$ (bottom panel of Fig.~2b) is significantly wider in the coupled case, showing a tenfold, or higher, probability of generating waves of 15m or larger. The coupling re-distributes the wave energy and introduce additional wave modes in the power spectra ${\cal F}\vert A \vert^2$, as a result of the coupling to the $B$-wave. Asymptotically (Fig.~2c), measured at $(\xi+2\zeta){\cal E} t \approx 55$ for the coupled system and $\xi {\cal E}t\approx 30$ for the uncoupled system, the differences are diminished, which also was indicated from the kurtosis in Fig~1c. Fig.~2d illustrates where the snapshots of Fig.~2a-c are taken.

\section{Discussion}

An increasing amount of data is currently being accumulated to improve rogue waves statistics and to map out areas prone to rogue waves~\cite{maxwave}. It is clear that such waves must be generated through a nonlinear mechanism. However, the treatment of coupled wave systems using a fully three-dimensional fluid dynamic setting is out of computational bounds. Therefore, although the model presented here does not take into account the full complexity of such wave interactions~\cite{dysthe}, it leaves the most essential nonlinear parts intact, and renders a statistical interpretation possible by performing large
scale simulations. Another interesting aspect of the model presented here is its suitability for optical experiments. Recently, the occurrence of optical rogue waves in white light propagation has been observed~\cite{solli}. Such direct optical experimental studies of extreme waves can be used as a means for discriminating between different waves models. Indeed, also here the propagation of light in optical fibers gives an simplified picture of extreme water surface wave dynamics, but in essence retain the important properties of such rogue wave formation. Measuring rouge wave statistics of different wave guides could thus be performed to experimentally investigate the effect of the wave coupling more closely. The simplest example would be to investigate the difference between reflective and absorbing boundaries in an optical fibre.

In conclusion, we observe two different dynamical scaling behaviors separated by a critical angle $\theta_c$---the direction of the wave systems relative to the dichotome. We also observe an increased probability of rogue wave occurrence in a coupled two-wave system compared with an uncoupled wave system. The increase is twofold; the interaction increases the nonlinear focusing of wave energy into larger waves, and additionally, the time needed to evolve large waves is decreased. The second part is particularly important when considering rogue waves at sea as wind speeds and directions vary significantly with time \cite{hasselmann,hasselmann80}. The local wind will align the waves in the wind direction, while turning winds are often associated with an inhomogeneous wind field making waves generated in different regions being radiated into neighboring regions. 

\acknowledgements
This work was partially supported by the Swedish Research Council (VR).

\end{document}